\documentclass[11pt]{article}

\PassOptionsToPackage{table}{xcolor}

\usepackage[final]{acl}

\usepackage[T1]{fontenc}
\usepackage[utf8]{inputenc}
\usepackage{times}
\usepackage{microtype}
\usepackage{graphicx}
\usepackage{float}
\usepackage{booktabs}
\usepackage{multirow}
\usepackage{amsmath}
\usepackage{amssymb}
\usepackage{pgfplots}
\pgfplotsset{compat=1.18}
\usepackage{enumitem}
\usepackage{fontawesome5}  
\definecolor{linkblue}{RGB}{31,56,180}  
\hypersetup{hidelinks}  
\usepackage{xspace}

\newcommand{\proposed}{proFILL\xspace}

\newcommand{\id}[1]{\texttt{#1}}
\newcommand{\gr}{\cellcolor[HTML]{D9D9D9}}

\title{hoBIT: A Profile-Aware Retrieval-Augmented Chatbot for
  University Academic Advising}

\author{
  Yoonseo Kim \\ Korea University \\ Seoul, Korea \\ {\small\texttt{seo3167@korea.ac.kr}}
  \And
  Seongmin Lee \\ Korea University \\ Seoul, Korea \\ {\small\texttt{kyne0127@korea.ac.kr}}
  \And
  Joongheon Kim \\ Korea University \\ Seoul, Korea \\ {\small\texttt{joongheon@korea.ac.kr}}
  \And
  SeongKu Kang\thanks{\,Corresponding author.} \\ Korea University \\ Seoul, Korea \\ {\small\texttt{seongkukang@korea.ac.kr}}
  \AND
  \normalfont\normalsize
  \href{https://github.com/hiyseo/hobit-emnlp}{\faGithub~\textcolor{linkblue}{Code}}\quad
  \href{https://youtu.be/IU4EcYIszZM}{\textcolor{red}{\faYoutube}~\textcolor{linkblue}{Demo Video}}\quad
  \href{https://idealab-ku.github.io/hobit-emnlp/}{\textcolor{linkblue}{\faGlobe~Homepage}}
}

\begin{document}
\maketitle

\begin{abstract}
In university academic advising, identical questions can require different answers depending on a student’s department, admission cohort, and degree program, causing profile-blind retrievers to surface plausible but inapplicable evidence.
We present \textbf{proFILL}, a method for transforming \textbf{hoBIT}, our college’s current rule-based advising chatbot, into a profile-aware retrieval-augmented generation (RAG) system.
Rather than requiring a complete user profile upfront, proFILL progressively acquires only the profile attributes needed for each query, guided by both the query intent and the initially retrieved evidence, and uses them to condition retrieval over a profile-aware index.
Extensive experiments and a human preference study show that \proposed outperforms diverse RAG baselines, is preferred by target users, and remains effective with open-weight models for cost-effective on-premise deployment.
\end{abstract}

\section{Introduction}
\label{sec:intro}

Retrieval-augmented generation (RAG) is increasingly employed in specialized domains \citep{gao2023ragsurvey}, including medical education \citep{jang2025medtutor} and religious question answering \citep{gao2025spiritrag}.
University academic advising is a practical and high-impact application.
Students frequently ask questions about graduation requirements, required courses, double-major rules, and scholarship eligibility, yet an incorrect answer can directly affect course planning, delay graduation, or increase the administrative workload of advising offices.

\begin{figure}[t]
\centering
\includegraphics[width=1.02\columnwidth]{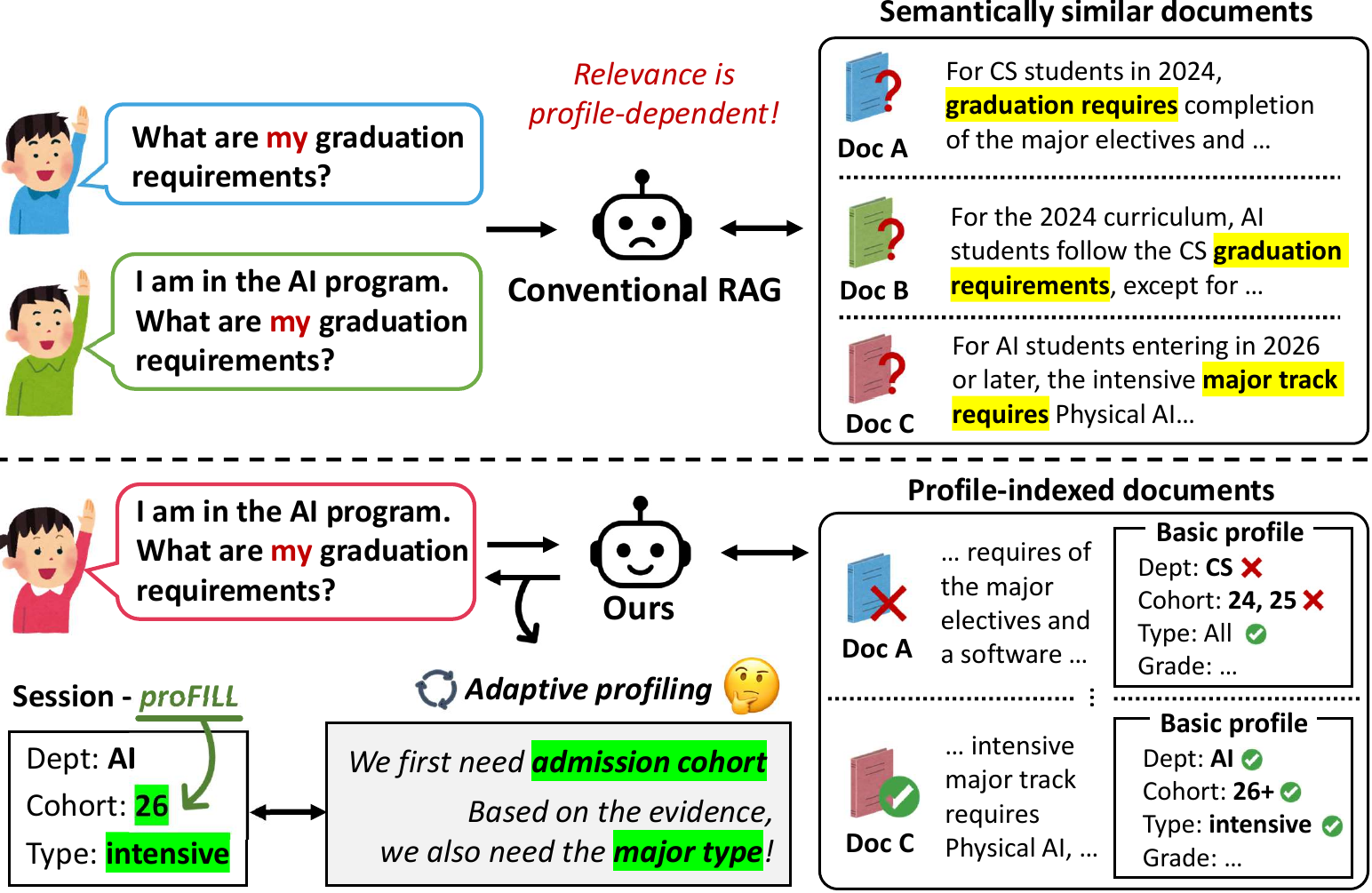}
\caption{A conceptual comparison of (top) conventional RAG, which struggles with profile-dependent queries over semantically similar documents, 
and (bottom) the proposed hoBIT, which adaptively acquires the required profile information and retrieves the applicable document from a profile-indexed corpus.}
\label{fig:motivation}
\end{figure}

The central difficulty is that academic advising questions are often \textit{profile-dependent}. 
The correct answer is not determined by the query alone, but also by a student’s profile, including department, admission cohort, major type, and related attributes.
In our setting, students belong to one of three departments: Computer Science (CS), Data Science (DS), and Artificial Intelligence (AI), where academic requirements and rules vary across admission cohorts.
Even within the same department, students may follow different rules depending on their major track, such as intensive, double-major, or convergence programs.
Thus, the same question may require different reference documents for different students.
Conventional RAG struggles in this setting, as semantically similar documents can lead a profile-blind retriever to return a plausible but wrong source (Figure~\ref{fig:motivation}).
Hand-written FAQ rules are also hard to maintain, as profile-specific cases accumulate with policy revisions over time.

To address this gap, we present \proposed, a method for transforming \textbf{hoBIT}, our college’s currently deployed rule-based academic advising chatbot, into a profile-aware retrieval-augmented system.
Our key idea is to treat profile-dependent evidence validity as part of the index structure, rather than as context to be resolved only at query time.
We implement this idea through \emph{offline profile-based indexing}, built on a lightweight academic profile schema covering department, admission cohort, major type, and related attributes.
During offline preprocessing, we use this schema to structure institutional materials, including regulations, department webpages, and notices.
Each chunk is annotated with the profile values for which it is valid and the profile fields needed to interpret it. 
This turns the corpus into profile-conditioned evidence, allowing retrieval to distinguish not only what a document is about, but also which students it applies to.

At online query time, we use this indexed structure to acquire the student profile on demand.
Instead of requiring login or a complete profile upfront, we maintain a time-limited session profile and collect only the fields needed by the current query.
This is achieved via \textit{on-demand adaptive-profiling}:
we first infer the query intent, identify the necessary profile fields, and ask only for missing ones.
The filled profile is combined with the query to match and filter against the indexed profile-conditioned evidence.
We then inspect the retrieved evidence and ask a follow-up question only when fine-grained additional information is needed to verify applicability, such as detailed scholarship eligibility conditions. 
Once the user provides the additional field, the session profile is updated and retrieval is rerun with the completed profile.
This evidence-driven selective process reduces initial user burden by avoiding a full upfront profile form, while improving answer correctness by triggering additional questions only when the retrieved evidence reveals unresolved profile requirements.

\noindent\textbf{Contributions.}
\textbf{(1)} We present \proposed, a method for transforming \textbf{hoBIT} into a profile-aware RAG system that replaces hand-written FAQ rules with profile-conditioned retrieval.
\textbf{(2)} In \proposed, we redesign both offline indexing and online query processing to address the profile-dependent nature of academic advising.
\textbf{(3)} Extensive experiments show that \proposed outperforms diverse RAG baselines, including profile-blind reranking and query-augmentation methods, while remaining effective across various open-weight models.

\section{The hoBIT System}
\label{sec:system}

\begin{figure*}[t]
\centering
\includegraphics[width=1.01\textwidth]{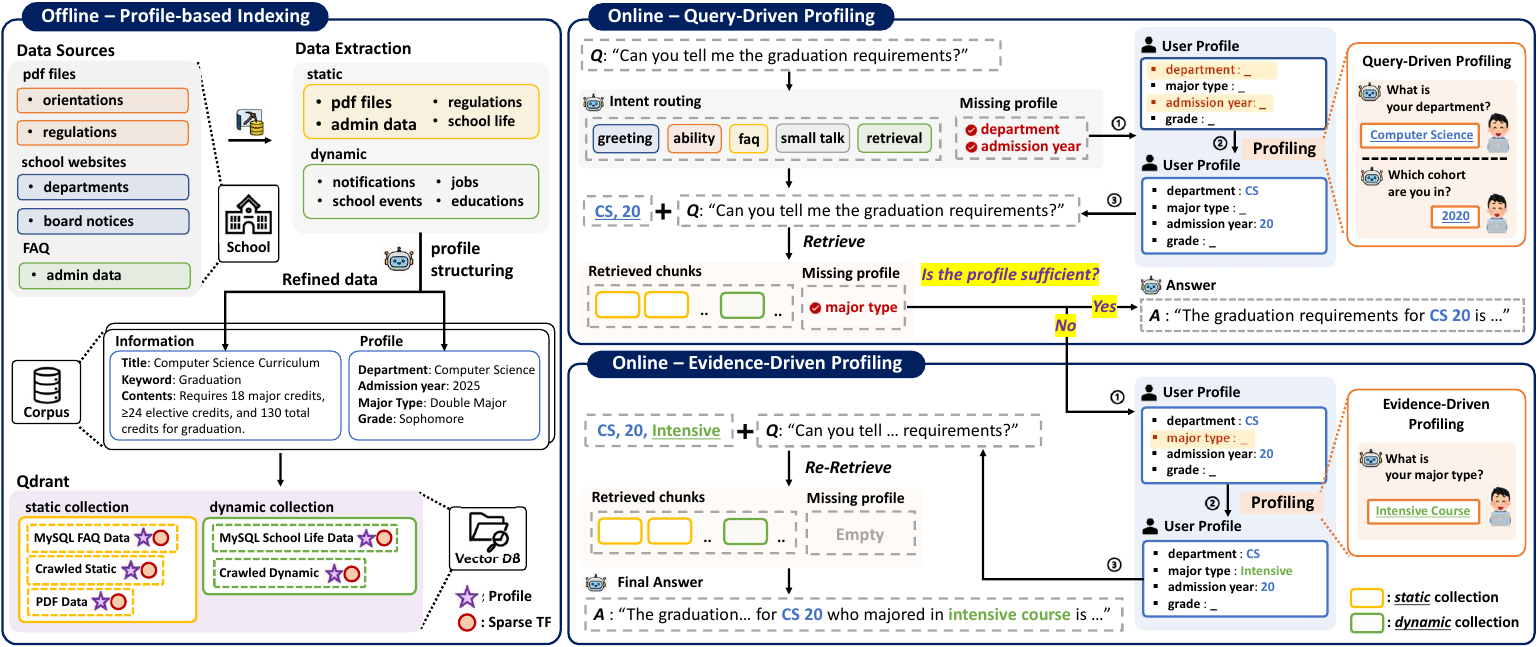}
\caption{Overview of hoBIT. \textbf{Offline (left):} each chunk is annotated with its applicable profile values and required fields before indexing. \textbf{Online (right):} query-driven profiling acquires missing attributes for profile-aware retrieval, while evidence-driven profiling requests additional information and re-retrieves when needed.}
\label{fig:pipeline}
\vspace{-1em}
\end{figure*}

\paragraph{Overview.}
hoBIT is an academic-advising chatbot deployed at our college of informatics.
Its original backend is a rule-driven dialogue system that maps user queries to predefined FAQ responses, limiting its ability to handle the profile-dependent questions common in academic advising.
We extend hoBIT with \proposed, a profile-aware retrieval-augmented generation framework featured with two key components: \emph{offline profile-aware indexing} and \emph{online on-demand profiling} (Figure~\ref{fig:pipeline}).

The resulting system can support a range of advising needs, including curriculum and graduation requirements, notices, and career-related questions.
The default profile schema comprises five attributes: \id{department}, \id{major type}, \id{grade}, \id{admission year} (i.e., admission cohort), and \id{student status}. 
These attributes determine which curricula, policies, and other institutional information apply.
The schema can be flexibly extended with additional attributes to support new advising needs.

\subsection{Offline Profile-based Indexing}
We build the corpus from five institutional sources: regulation PDFs, orientation PDFs, department webpages, board notices, and administrator-curated FAQs.
After cleaning and chunking the collected materials, we use an LLM to annotate each chunk with a structured profile record over the five predefined attributes.
Specifically, for each chunk, the LLM identifies which attributes determine the students to whom the information applies and fills in their corresponding values.
Attributes that do not affect the chunk’s applicability are left \id{null}.
The resulting non-null fields encode the chunk’s profile-dependent applicability and unveil which user attributes must be known for reliable retrieval.

In practice, we maintain separate static and dynamic indices based on source update frequency.
Relatively stable materials (e.g., regulations) are stored in the static index, whereas frequently updated materials (e.g., notices) are stored in the dynamic index.
Results from both indices are combined at query time.
Further details on preprocessing and indexing are provided in Appendix~\ref{sec:app-pipeline}.

\subsection{Online Query Processing}

\subsubsection{Intent Routing and Session-Profile Initialization}

\noindent \textbf{Intent Routing.}
At query time, an LLM first classifies each incoming query into one of five intents: \emph{greeting}, \emph{ability}, \emph{faq}, \emph{smalltalk}, and \emph{retrieval}.
Only \emph{retrieval} queries proceed to subsequent retrieval and answer generation.
\emph{greeting}, \emph{ability}, and \emph{faq} queries are handled using predefined templates or the FAQ store without an additional LLM generation call, while \emph{smalltalk} queries receive a lightweight conversational response.

\vspace{5pt}
\noindent \textbf{Session-Profile Initialization.}
For each \emph{retrieval} query, the LLM identifies the profile attributes required to answer the question and extracts any corresponding values available from the query.
These values initialize the session profile, while required attributes whose values are unavailable are marked as missing.
These missing values are subsequently acquired as needed via adaptive profiling.

Importantly, the initialized profile serves only as a guide rather than a fixed specification of all information required to answer the query.
Additional information that cannot be inferred from the query or is not covered by the predefined profile schema may be acquired on demand when its necessity becomes evident from the retrieved evidence.

\subsubsection{On-demand Adaptive Profiling and Profile-Aware RAG}
Rather than requiring login or a complete profile form upfront, \proposed acquires only the information needed for the current query through two complementary stages: \emph{query-driven profiling} and \emph{evidence-driven profiling}.

\paragraph{Query-Driven Profiling and Retrieval.}
Before retrieval, the system asks the user for the attributes marked as missing during session-profile initialization.
The acquired values are added to the session profile and incorporated into retrieval through both \textit{soft} query augmentation and \textit{hard} filtering.
Specifically, the available profile values are serialized and prepended to the original query for retrieval.\footnote{Further details on retrieval are provided in Appendix~\ref{sec:app-retrieval}.}
The retrieved candidates are then filtered using the profile annotations: 
Chunks whose specified profile values conflict with the session profile are removed, while chunks with matching or unspecified (null) profile values remain eligible.
The top-10 eligible chunks are provided to the generator LLM.

Before generating an answer, an LLM selects the chunks needed to answer the query as the evidence set and checks whether their applicability can be determined from the current session profile.
Specifically, it identifies any non-null profile fields of the selected chunks whose corresponding attributes remain unresolved, as these fields indicate the information required to determine whether each chunk applies to the user.
The LLM also checks whether additional user information is needed to interpret the evidence reliably.
If no further information is required, the system generates an answer from the selected evidence; otherwise, evidence-driven profiling is triggered.\footnote{In our setup, this occurs for 24.3\% of queries on average.}

\paragraph{Evidence-Driven Profiling and Re-Retrieval.}
When additional information is required, the system asks the user a targeted follow-up question about the unresolved profile fields or other required information.
The acquired information is added to the session profile, and retrieval is repeated in the same manner using the updated profile.
An LLM then reselects the relevant evidence and generates the final answer from the refined evidence set.

\subsection{System Implementation}
To improve usability and user trust, hoBIT incorporates two interaction design features.
First, for attributes covered by the predefined profile schema (e.g., \id{admission year}), the system presents predefined options instead of requiring free-form input, reducing user effort.
Second, each generated answer is accompanied by the selected evidence, allowing users to directly inspect the institutional sources supporting the response.

\section{Experimental Setup}
\label{sec:setup}

\paragraph{Corpus.}
We construct the corpus from 515 institutional sources collected from our college: 3 academic-regulation and orientation PDFs, 89 department webpages, 137 board notices, and 286 administrator-maintained FAQ entries.

\paragraph{Profile-Grounded QA Data.}
With guidance from our college's academic affairs office and drawing on query logs from the deployed hoBIT service, we use LLM-assisted generation to construct 1,800 QA instances spanning 60 unique student profiles, 10 advising categories,\footnote{The categories cover major requirements, general education requirements, core general education, foundational courses, graduation requirements and projects, broad versus intensive majors, general education areas, department-specific major electives, graduation-credit breakdowns, and credit recognition for double majors.} and three query types: formal, first-person, and affirmative or negative verification queries.
Further details on dataset construction and additional experiments for a broader evaluation, including intent routing and open-ended advising, are presented in Appendix~\ref{sec:appendix-aux}.

\paragraph{Evaluation Settings.}
We evaluate the profile-grounded QA under two settings.
\textbf{Deployment} reflects the realistic scenario in which the user profile is initially unavailable and must be acquired during interaction.
\textbf{Oracle} provides the complete profile along with the query, allowing us to assess how effectively \proposed leverages the available profile information.
For RAG baselines, the given profile is linearized and appended to the query as additional context.
For all methods, we use \texttt{text-embedding-3-small} as the dense retriever and \texttt{gpt-4o-mini} as the generator.
Latency is measured using a 48 GB MIG instance on a single NVIDIA RTX PRO 6000 GPU and an Intel Xeon Gold 6530 CPU.

\paragraph{Metrics.}
We evaluate retrieval using MRR and Recall@$\{1,5,10,50\}$.
Generation is evaluated using three metric types: 
(1) lexical metrics (ROUGE-L and Token-F1), which measure surface-level overlap with reference answers;
(2) matching metrics (Keyword Match and Source Match), which assess whether the answer includes the required key information and uses the correct sources; and 
(3) the LLM-based metric (Grounded Correctness), which evaluates whether the answer is both correct and grounded in appropriate sources.\footnote{We use \texttt{deepeval}~\citep{deepeval} and report scores averaged across \texttt{Qwen2.5-32B-Instruct} and \texttt{gpt-4o-mini}. 
}
Further details are provided in Appendix~\ref{sec:app-metrics}.
\section{Results and Discussion}
\label{sec:results-new}

\begin{table*}[t!]
\centering\footnotesize
\setlength{\tabcolsep}{5pt}
\resizebox{\textwidth}{!}{%
\begin{tabular}{c | l ccccc ccccc |c}
\toprule
\multicolumn{1}{c}{} & & \multicolumn{5}{c}{\textbf{Retrieval}} & \multicolumn{5}{c}{\textbf{Generation}} & \\
\cmidrule(lr){3-7}\cmidrule(lr){8-12}
\multicolumn{1}{c}{} & \textbf{Retrieval System} & MRR & R@1 & R@5 & R@10 & R@50 & ROUGE-L & Token-F1 & Keyword Match & Source Match & Grounded Correctness & s/q \\
\midrule
\multirow{8}{*}{\rotatebox[origin=c]{90}{\emph{Deployment}}}
 & BM25 & 0.089 & 0.017 & 0.118 & 0.267 & 0.728 & 0.139 & 0.167 & 0.564 & 0.412 & 0.412 & 4.0 \\
 & Dense & 0.031 & 0.011 & 0.022 & 0.049 & 0.400 & 0.149 & 0.176 & 0.548 & 0.217 & 0.292 & 4.3 \\
 & Hybrid & 0.061 & 0.009 & 0.069 & 0.152 & 0.698 & 0.146 & 0.173 & 0.574 & 0.380 & 0.395 & 4.3 \\
 & \quad $+$reranker & 0.111 & 0.027 & 0.165 & 0.322 & 0.698 & 0.152 & 0.177 & 0.621 & 0.391 & 0.425 & 4.6 \\
 & HyDE & 0.061 & 0.013 & 0.057 & 0.141 & 0.689 & 0.147 & 0.175 & 0.586 & 0.367 & 0.403 & 6.9 \\
 & \quad $+$reranker & 0.103 & 0.025 & 0.156 & 0.300 & 0.690 & 0.151 & 0.174 & 0.613 & 0.379 & 0.422 & 7.2 \\
 & \gr proFILL & \gr \textbf{0.593} & \gr \textbf{0.492} & \gr \textbf{0.718} & \gr \textbf{0.782} & \gr \textbf{0.936} & \gr \textbf{0.172} & \gr \textbf{0.199} & \gr \textbf{0.699} & \gr \textbf{0.749} & \gr \textbf{0.625} & \gr 6.2 \\
 & \gr \quad $+$reranker & \gr 0.420 & \gr 0.271 & \gr 0.561 & \gr 0.691 & \gr \textbf{0.936} & \gr 0.168 & \gr 0.193 & \gr 0.665 & \gr 0.694 & \gr 0.596 & \gr 6.4 \\
\midrule[\heavyrulewidth]
\multirow{8}{*}{\rotatebox[origin=c]{90}{\emph{Oracle}}}
 & BM25 & 0.211 & 0.059 & 0.329 & 0.624 & 0.979 & 0.155 & 0.184 & 0.648 & 0.646 & 0.555 & 4.0 \\
 & Dense & 0.475 & 0.306 & 0.693 & 0.871 & 0.994 & 0.171 & 0.200 & 0.725 & 0.723 & 0.617 & 4.3 \\
 & Hybrid & 0.401 & 0.219 & 0.647 & 0.841 & 0.993 & 0.171 & 0.199 & 0.716 & 0.705 & 0.621 & 4.4 \\
 & \quad $+$reranker & 0.151 & 0.033 & 0.224 & 0.444 & 0.993 & 0.160 & 0.186 & 0.623 & 0.583 & 0.530 & 4.5 \\
 & HyDE & 0.424 & 0.242 & 0.683 & 0.886 & 0.992 & 0.169 & 0.198 & 0.716 & 0.695 & 0.608 & 7.1 \\
 & \quad $+$reranker & 0.159 & 0.039 & 0.229 & 0.487 & 0.993 & 0.164 & 0.191 & 0.643 & 0.591 & 0.538 & 7.2 \\
 & \gr proFILL & \gr \textbf{0.780} & \gr \textbf{0.674} & \gr \textbf{0.929} & \gr \textbf{0.980} & \gr \textbf{1.000} & \gr \textbf{0.179} & \gr \textbf{0.207} & \gr \textbf{0.733} & \gr \textbf{0.847} & \gr \textbf{0.676} & \gr 4.4 \\
 & \gr \quad $+$reranker & \gr 0.549 & \gr 0.397 & \gr 0.704 & \gr 0.829 & \gr \textbf{1.000} & \gr 0.174 & \gr 0.201 & \gr 0.685 & \gr 0.787 & \gr 0.638 & \gr 4.5 \\
\bottomrule
\end{tabular}}
\caption{Retrieval and generation performance. For all methods, \texttt{gpt-4o-mini} is used as the answer generator. \emph{s/q} denotes end-to-end latency in seconds per query; absolute latency may vary with the hardware configuration and external LLM API response time.}
\label{tab:main}
\end{table*}

\paragraph{Comparison with RAG Baselines.}
Table~1 compares \proposed with RAG baselines using the same generator but different retrieval strategies.
Hybrid combines BM25 and dense retrieval, HyDE~\citep{gao2022hyde} augments the query with LLM-generated context, and Reranker applies a cross-encoder to refine the retrieved results.\footnote{We use \texttt{BAAI/bge-reranker-v2-m3}.}

Overall, \proposed achieves the strongest retrieval and generation performance.
In the \emph{deployment} setting, it outperforms all baselines in MRR, including those given complete profiles under the \emph{oracle} setting ($0.593$ vs.\ $0.475$).
It also shows clear gains in lexical and matching metrics, including Source Match ($0.749$ vs.\ $0.412$), and achieves the highest Grounded Correctness in both settings.
Interestingly, adding a profile-blind reranker to \proposed degrades performance, indicating that reranking without profile information can be harmful once profile applicability has already been incorporated into retrieval.
\proposed also remains efficient, requiring $6.2$ seconds per query, compared with $6.9$ seconds for HyDE.
The $1.8$-second gap from oracle \proposed ($4.4$ seconds) reflects the additional cost of on-demand adaptive profiling.

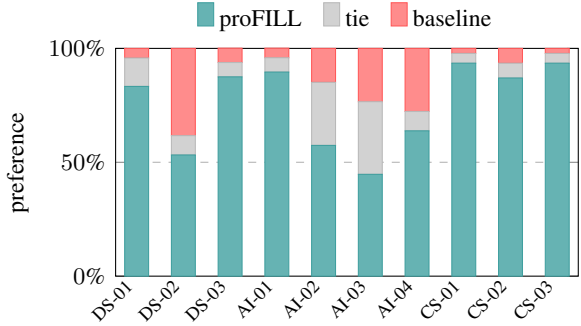
\begin{figure}[t]
\centering
\begin{tikzpicture}
\begin{axis}[
  ybar stacked, width=\linewidth, height=4.6cm, bar width=9pt,
  ymin=0, ymax=100, ytick={0,50,100},
  yticklabel={\pgfmathprintnumber[precision=0]{\tick}\%}, yticklabel style={font=\footnotesize},
  ylabel={preference}, ylabel style={font=\footnotesize},
  symbolic x coords={DS-01,DS-02,DS-03,AI-01,AI-02,AI-03,AI-04,CS-01,CS-02,CS-03},
  xtick=data, x tick label style={font=\scriptsize, rotate=45, anchor=east},
  enlarge x limits=0.05, clip=false,
  legend style={at={(0.5,1.03)}, anchor=south, legend columns=-1, font=\footnotesize, draw=none, fill=none,
    /tikz/every even column/.append style={column sep=6pt}},
]
\addplot[fill=teal!60,draw=teal!75] coordinates {(DS-01,83.3)(DS-02,53.2)(DS-03,87.5)(AI-01,89.6)(AI-02,57.4)(AI-03,44.7)(AI-04,63.8)(CS-01,93.5)(CS-02,87.0)(CS-03,93.5)};
\addlegendentry{proFILL}
\addplot[fill=gray!35,draw=gray!55] coordinates {(DS-01,12.5)(DS-02,8.5)(DS-03,6.3)(AI-01,6.3)(AI-02,27.7)(AI-03,31.9)(AI-04,8.5)(CS-01,4.3)(CS-02,6.5)(CS-03,4.3)};
\addlegendentry{tie}
\addplot[fill=red!45,draw=red!65] coordinates {(DS-01,4.2)(DS-02,38.3)(DS-03,6.3)(AI-01,4.2)(AI-02,14.9)(AI-03,23.4)(AI-04,27.7)(CS-01,2.2)(CS-02,6.5)(CS-03,2.2)};
\addlegendentry{baseline}
\draw[dashed,gray!60] (axis cs:DS-01,50)--(axis cs:CS-03,50);
\end{axis}
\end{tikzpicture}
\caption{Human preference comparison. Results are reported as three-way vote splits. Question IDs are prefixed by the corresponding department.}
\label{fig:userstudy}
\end{figure}

\paragraph{Human Preference Evaluation.}
To evaluate user preference within hoBIT's actual target population, we conduct a blind pairwise study with 48 students from our college.
Participants compare \proposed against conventional dense retrieval-based RAG on ten profile-dependent questions (Appendix~\ref{sec:app-userstudy}). 
In Figure~\ref{fig:userstudy}, \proposed is preferred on all ten questions, achieving an aggregate non-tie win rate of $85.3\%$ (354 wins vs.\ 61 losses; two-sided binomial test, $p<0.001$).
Preference is highest for graduation- and credit-related questions ($93$--$98\%$; DS-01, AI-01, CS-02, CS-03), whose answers strongly depend on the student's profile, and lowest for the broader question on available major courses ($58\%$; DS-02).

\begin{table}[t]
\centering\footnotesize
\setlength{\tabcolsep}{4pt}
\renewcommand{\arraystretch}{0.95}
\resizebox{\columnwidth}{!}{%
\begin{tabular}{l | l cccc c}
\toprule
\multicolumn{1}{l}{\textbf{Embedder}} & \textbf{Setting} & MRR & R@1 & R@5 & R@10 & R@50 \\
\midrule
\multirow{4}{*}{\id{openai\_small}}
 & \gr proFILL & \gr\textbf{0.596} & \gr\textbf{0.495} & \gr\textbf{0.721} & \gr\textbf{0.784} & \gr\textbf{0.936} \\
 & \quad w/o Hard filtering        & 0.311 & 0.157 & 0.502 & 0.671 & 0.923 \\
 & \quad w/o Soft aug.       & 0.247 & 0.137 & 0.337 & 0.501 & 0.893 \\
  & \quad w/o Re-retrieval   & 0.580 & 0.481 & 0.702 & 0.767 & 0.934 \\
\midrule
\multirow{4}{*}{\id{BGE-M3}}
 & \gr proFILL & \gr\textbf{0.665} & \gr\textbf{0.579} & \gr\textbf{0.775} & \gr\textbf{0.796} & \gr0.936 \\
 & \quad w/o Hard filtering        & 0.475 & 0.349 & 0.624 & 0.704 & 0.934 \\
 & \quad w/o Soft aug.        & 0.347 & 0.197 & 0.517 & 0.709 & \textbf{0.938} \\
  & \quad w/o Re-retrieval    & 0.644 & 0.558 & 0.750 & 0.776 & 0.934 \\
\midrule
\multirow{4}{*}{\id{Qwen3-Emb}}
 & \gr proFILL & \gr\textbf{0.651} & \gr\textbf{0.564} & \gr\textbf{0.763} & \gr\textbf{0.801} & \gr\textbf{0.931} \\
 & \quad w/o Hard filtering        & 0.425 & 0.304 & 0.574 & 0.708 & 0.926 \\
 & \quad w/o Soft aug.        & 0.304 & 0.184 & 0.408 & 0.556 & 0.919 \\
 & \quad w/o Re-retrieval    & 0.636 & 0.549 & 0.748 & 0.785 & 0.927 \\
\bottomrule
\end{tabular}}
\caption{Ablation study. Results are reported with three different dense embedding models for retrieval: \texttt{text-embedding-3-small}~\citep{text_embedding_3}, BGE-M3~\citep{chen2024bgem3}, and Qwen3-Embedding~\citep{qwen3emb}.}
\label{tab:retriever}
\end{table}

\paragraph{Ablation Study.}
Table~2 reports ablation results for \proposed.
Both profile-injection mechanisms are critical: removing either the soft prefix or hard filter substantially degrades retrieval performance.
Disabling re-retrieval after evidence-driven profiling also consistently degrades performance.
Notably, Recall@50 remains near ceiling without re-retrieval, indicating that its primary benefit is to promote relevant sources to higher ranks, allowing the generator to access necessary evidence with fewer chunks and use its context window more efficiently.
These trends are consistent across all three dense embedders.

\paragraph{Results with Varying LLMs.}
In Table~\ref{tab:generator}, we evaluate \proposed with various LLM generators.
We compare \texttt{gpt-4o-mini} with three multilingual open-weight models (\texttt{Qwen3-8B}, \texttt{Ministral-8B}, and \texttt{LLaMA3.1-8B}) and two Korean-specialized models (\texttt{EXAONE3.5-7.8B} and \texttt{Kanana1.5-8B}).
Open-weight models remain competitive with \texttt{gpt-4o-mini}; notably, the Korean-specialized models perform strongly on matching metrics, suggesting their suitability for grounded advising in Korean-language settings.

\begin{table}[t]
\centering
\setlength{\tabcolsep}{4pt}
\renewcommand{\arraystretch}{0.95}
\resizebox{\columnwidth}{!}{%
\begin{tabular}{c | l cc cc c}
\toprule
\multicolumn{1}{c}{} & & \multicolumn{2}{c}{\emph{Lexical}} & \multicolumn{2}{c}{\emph{Matching}} & \\
\cmidrule(lr){3-4}\cmidrule(lr){5-6}
\multicolumn{1}{c}{} & \textbf{Generator} & ROUGE-L & Token-F1 & Kw.\ Match & Src.\ Match & GC \\
\midrule
\emph{Prop.}
  & \texttt{gpt-4o-mini} & 0.179 & 0.207 & 0.733 & 0.847 & 0.676 \\
\midrule
\multirow{5}{*}{\shortstack[l]{\emph{Open-}\\\emph{weight}}}
  & \texttt{Qwen3-8B} & 0.169 & 0.197 & 0.721 & 0.879 & 0.670 \\
  & \texttt{Ministral-8B} & 0.155 & 0.177 & 0.723 & 0.734 & 0.639 \\
  & \texttt{LLaMA3.1-8B} & \textbf{0.204} & \textbf{0.230} & 0.665 & 0.817 & 0.649 \\
  & \texttt{EXAONE3.5-7.8B} & 0.135 & 0.158 & 0.704 & \textbf{0.898} & \textbf{0.712} \\
  & \texttt{Kanana1.5-8B} & 0.123 & 0.143 & \textbf{0.752} & 0.861 & 0.697 \\
\bottomrule
\end{tabular}}
\caption{Results across various LLMs, including proprietary and open-weight models. GC denotes Grounded Correctness.}
\label{tab:generator}
\end{table}

\begin{figure*}[t]
\centering
\includegraphics[width=\textwidth]{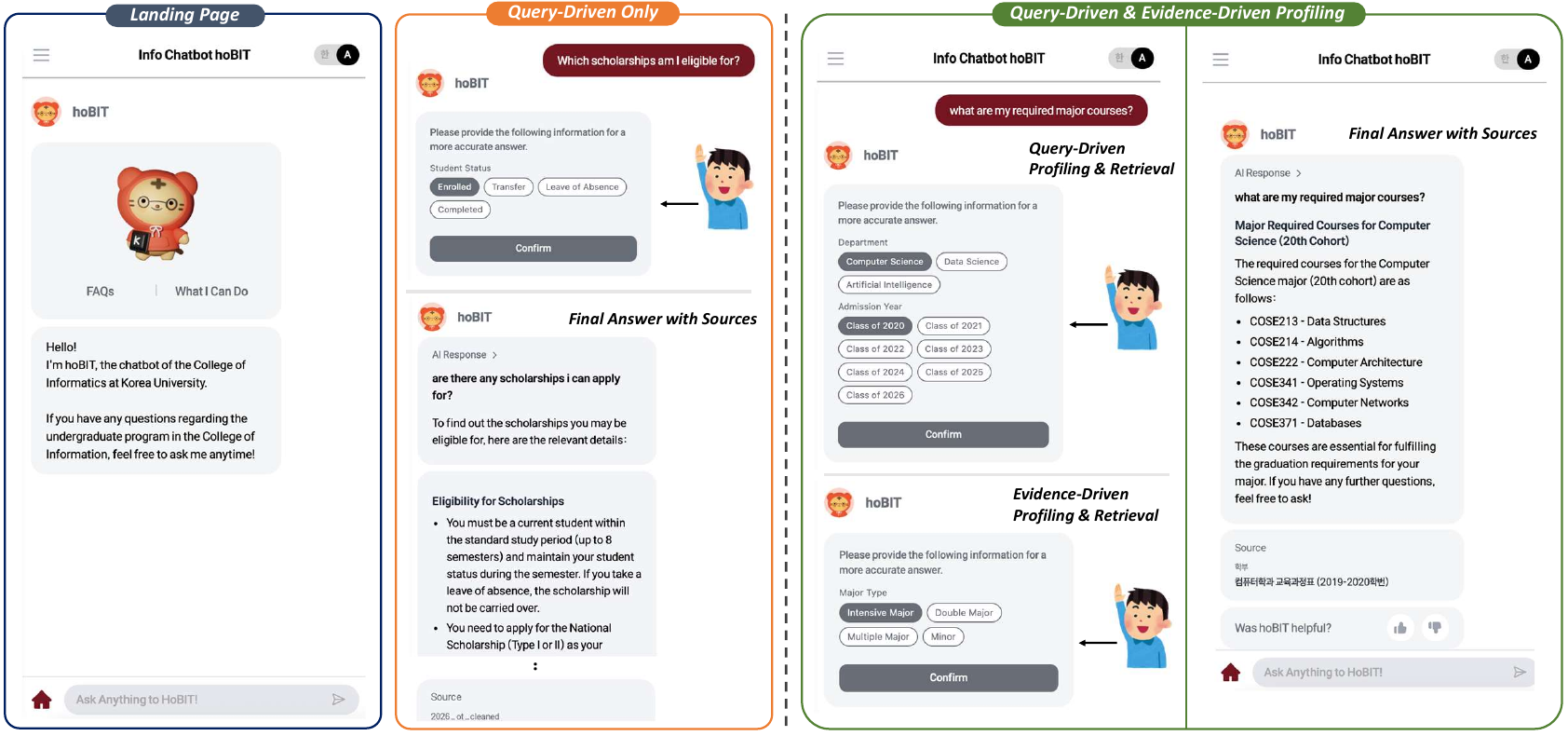}
\caption{The web interface with examples for two queries. \proposed requests only the information needed for each query on demand and uses it to answer the user’s question.}
\label{fig:ui}
\end{figure*}

\section{System Demonstration}
\label{sec:demo}

Figure~\ref{fig:ui} illustrates two separate interactions with hoBIT using \proposed.
For the first scholarship query, query-driven profiling obtains the student’s status.
As this information is sufficient to answer the query, no additional profiling is triggered, and the system directly provides an answer with supporting~sources.

For the second major-course query, \proposed first collects the department and admission cohort.
The initially retrieved evidence then indicates that the applicable curriculum depends on the student’s major type, triggering an additional evidence-driven profiling step.
After obtaining this information, \proposed re-retrieves the relevant documents and returns a profile-specific answer with supporting sources.

\section{Related Work}
\label{sec:related}

\noindent\textbf{Retrieval-augmented generation.}
RAG~\citep{lewis2020rag} grounds generation in retrieved evidence using dense~\citep{karpukhin2020dpr}, lexical~\citep{bm25}, or hybrid retrieval~\citep{rrf2009}, and can be improved through reranking~\citep{nogueira2019rerank}, query augmentation~\citep{gao2022hyde}, and structured indexing~\citep{microsoft_graphrag2024}.
\proposed instead conditions retrieval on an explicit user profile.

\noindent\textbf{Personalization in LLMs.}
Prior work personalizes LLMs using user preferences~\citep{choi2025copl,thonet2025fast}, interaction histories~\citep{qin2025vaps,li2025multisession,su2025pqa}, or personas~\citep{persona_rag}.
In contrast, \proposed injects an explicit schema-typed profile into retrieval through soft query augmentation and hard metadata filtering, without per-user training.

\noindent\textbf{Academic-advising chatbots.}
University assistants have evolved from intent-based dialogue to retrieval-grounded systems.
The closest system, Marcel~\citep{trienes2025marcel}, answers questions from university resources but does not explicitly condition retrieval on a student profile.
\proposed instead uses adaptive profiling to retrieve cohort-specific evidence and tailor answers to each student.
\section{Conclusion}
\label{sec:conclusion}
We present \proposed, which transforms hoBIT from a rule-based chatbot into a profile-aware RAG system through offline profile indexing and on-demand adaptive profiling.
It acquires missing information through query-driven profiling and requests additional details through evidence-driven profiling when needed.
Experiments on institutional data provide practical insights into how structured profiles improve RAG for academic~advising.

\section*{Limitations}
First, the corpus and benchmark are drawn from a single college of informatics. 
Although \proposed's overall pipeline is domain-general, its profile schema must be adapted to the curricula, policies, and advising practices of each educational institution.
Second, \proposed relies on self-reported profile information without institutional verification, so incorrect inputs (e.g., an inaccurate admission year) may propagate to the retrieval results.

\section*{Ethics Statement}
\paragraph{Data and Privacy.}
By design, \proposed acquires only the profile attributes required for the current query, retains them only for the current session, and discards them afterward; no persistent per-user profile database is maintained.
The usage logs used in this work contain no personally identifiable information and serve only to validate the query distribution---never to generate benchmark items, whose student profiles are entirely synthetic.
The human preference study was voluntary and recorded only participants' department affiliations and pairwise preferences.

\paragraph{Responsible Use.}
hoBIT is an assistive information tool rather than an authoritative source. 
Every answer includes citations to institutional sources for verification, and any binding decision remains subject to the university's official regulations.

\section*{Acknowledgments}
We thank the Korea University College of Informatics and the KU Web Development Club.
This work was supported by the IITP grant funded by the MSIT (IITP-2026-RS-2020-II201819), the NRF grant funded by the MSIT (RS-2026-25486220), and the MSIT under the ITRC program (IITP-2026-RS-2024-00436887), supervised by the IITP.

\bibliography{hobit-refs}

\clearpage
\appendix
\section*{Appendix}
All system and LLM-judge prompts, the dataset, and the human evaluation questionnaire are available in our code repository (\href{https://github.com/hiyseo/hobit-emnlp}{\textcolor{blue}{\underline{link}}}) and on the project website (\href{https://idealab-ku.github.io/hobit-emnlp/}{\textcolor{blue}{\underline{link}}}).

\section{Offline Indexing}
\label{sec:app-pipeline}
Each cleaned document is segmented into self-contained semantic chunks of up to 800 characters using paragraph-level splitting followed by sentence-level splitting.
Before indexing, each chunk is prepended with its topic, subtopic, title, and extracted keywords to improve lexical matching.
Expired notices are removed, and administrator-FAQ entries are consolidated.
An LLM tagger (\texttt{gpt-4o-mini}) annotates each chunk over the five predefined profile attributes.
For each attribute, it assigns the value or range of students to whom the chunk applies and sets unrelated attributes to \id{null}.
The resulting non-null fields explicitly encode the chunk's profile-dependent applicability.

Chunks are divided by update frequency into a \emph{static} collection for stable content and a \emph{dynamic} collection for frequently updated notices.
Each collection maintains sparse and dense indices for hybrid retrieval.
The sparse index uses Kiwi-based Korean tokenization with content-morpheme filtering and 32K-dimensional feature hashing for BM25 scoring.
The dense index uses \texttt{text-embedding-3-small} vectors.

\section{Retrieval Process}
\label{sec:app-retrieval}
Given a query, we first translate non-Korean input into Korean.
It then performs hybrid retrieval independently over the static and dynamic collections and combines their results through time-aware aggregation to construct the final retrieval results.

\paragraph{Hybrid Retrieval.}
Each collection is searched using both dense and BM25 retrieval, capturing overall semantic relevance and lexical term matching, respectively.
The two rankings are combined using reciprocal rank fusion with $k{=}60$.

\paragraph{Time-Aware Aggregation.}
The system internally determines whether a query is time-sensitive and selects ten chunks accordingly.
For time-insensitive queries, it selects seven results from the static collection and three from the dynamic collection; this allocation is reversed for time-sensitive queries.
Within the dynamic collection, a recency score with a 90-day half-life favors newer notices among similarly relevant results.

\section{Dataset Construction and Supplementary Results}
\label{sec:appendix-aux}
All datasets were constructed under the guidance of our college’s academic affairs office and using query logs from the deployed hoBIT service, with \texttt{GPT-4o-mini} employed for LLM-assisted generation.
No personally identifiable information was collected during dataset construction.

\paragraph{Profile-grounded QA.}
The index contains 906 chunks segmented and profile-annotated from the collected institutional sources (\S\ref{sec:system}).
The main dataset is constructed solely from the static collection, and retrieval during evaluation is restricted to the same collection.
It comprises 1,800 QA instances, covering all combinations of 60 student profiles, 10 profile-dependent advising categories, and three query types: formal, first-person, and verification-style.
The 60 profiles combine 15 department--admission-year cohorts across CS, DS, and AI with four grade levels.
Admission cohorts are further grouped into eight curriculum-revision periods, which determine the applicable curriculum for each profile.

Each instance includes a deterministic gold source and expected keywords validated against the index.
Profile information is provided only through the session profile and is excluded from the query text, isolating the effect of profile-aware retrieval.
The advising categories were curated from the indexed materials and cross-checked against 797 query anchors distilled from 3,058 historical hoBIT logs, covering 14 of the 15 observed log categories.

\paragraph{Intent Routing.}
We construct an intent-routing dataset of 1,600 queries covering five intents: \emph{greeting}, \emph{ability}, \emph{faq}, \emph{smalltalk} and \emph{retrieval}.
The dataset is assembled from three sources.
First, we manually label 22 queries from real hoBIT service logs to obtain seed examples for the four non-retrieval intents.
Second, using these seeds, we generate 378 additional queries using the LLM, resulting in 100 queries for each non-retrieval intent and 400 queries in total.
Third, we reuse the 1,200 queries from the open-ended advising dataset as \emph{retrieval} queries, ensuring that the routing and downstream RAG evaluations follow the same domain distribution.
Table~\ref{tab:intent} shows that the proposed framework accurately identifies query intents, achieving an F1 score of $0.990$ for \emph{retrieval} queries.

\begin{table}[t]
\centering\footnotesize
\setlength{\tabcolsep}{5pt}
\renewcommand{\arraystretch}{0.95}
\begin{tabular}{l ccc c}
\toprule
\textbf{Intent class} & Precision & Recall & F1 & \#Queries \\
\midrule
greeting  & 1.000 & 0.630 & 0.773 & 100 \\
ability   & 0.842 & 0.960 & 0.897 & 100 \\
faq       & 0.971 & 1.000 & 0.985 & 100 \\
smalltalk & 0.704 & 1.000 & 0.826 & 100 \\
retrieval  & 1.000 & 0.981 & 0.990 & 1{,}200 \\
\bottomrule
\end{tabular}
\caption{Intent classification results.}
\label{tab:intent}
\end{table}

\paragraph{Open-ended Advising.}
We additionally evaluate open-ended advising questions spanning 12 academic and student-life categories.
Since these questions have no deterministic reference answers, we use LLM judges to evaluate two metrics (Table~\ref{tab:openended}).
\emph{Top-3 Precision} measures the proportion of the top-3 retrieved chunks judged relevant to the question.
\emph{Answer Completeness}, implemented with \texttt{deepeval}, measures how fully the answer resolves the question given the retrieved context.
Answers that fully use the available evidence score highly, whereas evasive or hallucinated responses score poorly. An explicit ``information not found'' response is rewarded when the retrieved context genuinely lacks the answer.

\begin{table}[t]
\centering\footnotesize
\setlength{\tabcolsep}{4pt}
\renewcommand{\arraystretch}{0.95}
\resizebox{\columnwidth}{!}{%
\begin{tabular}{l cc c}
\toprule
\textbf{Category} & Top-3 Precision & Completeness & \#Queries \\
\midrule
Academic status      & 0.713 & 0.973 & 100 \\
Facilities/dining    & 0.683 & 0.931 & 100 \\
Academic operations  & 0.670 & 0.935 & 100 \\
Scholarships         & 0.643 & 0.933 & 100 \\
Space/facility       & 0.633 & 0.944 & 100 \\
Enrollment/tuition   & 0.593 & 0.921 & 100 \\
Clubs/council        & 0.610 & 0.909 & 100 \\
Course registration  & 0.593 & 0.926 & 100 \\
Student services     & 0.567 & 0.935 & 100 \\
Graduate/research    & 0.610 & 0.907 & 100 \\
Notices/schedule     & 0.403 & 0.871 & 100 \\
Career/internship    & 0.279 & 0.880 & 99 \\
\midrule
\textbf{Overall}     & 0.584 & 0.922 & 1{,}199 \\
\bottomrule
\end{tabular}}
\caption{Open-ended advising results by category.}
\label{tab:openended}
\end{table}

\section{Evaluation Metrics and LLM Judges}
\label{sec:app-metrics}

\paragraph{Lexical and Matching Metrics.}
\emph{ROUGE-L} and \emph{Token-F1} measure overlap with reference answers after both texts are tokenized into Korean content morphemes using Kiwi, reducing penalties from inflectional variation.
\emph{Keyword Match} measures the proportion of expected content keywords included in the answer.
\emph{Source Match} evaluates attribution over the static and dynamic collections: a cited gold source scores 1, one retrieved but not cited scores 0.5, and a missing source scores 0, thereby rewarding grounded citation beyond retrieval alone.

\paragraph{LLM-based Metrics.}
We evaluate generation quality using \emph{Grounded Correctness} (GC), which combines LLM-judged \emph{Answer Correctness} (AC) and \emph{Source Match} (SM):
\[
\mathrm{GC}=\sqrt{\mathrm{AC}\times\mathrm{SM}}.
\]
The geometric mean rewards answers only when they are both correct and grounded in the appropriate evidence.
\emph{Answer Correctness} is implemented with \texttt{deepeval}: questions are classified as \emph{positive-verification}, \emph{negative-verification}, or \emph{general}; verification questions evaluate only the requested yes/no decision, while general questions assess coverage of the expected key items.
Scores are averaged across two independent judges on a 600-case subset stratified by the 10 categories and 3 phrasing types (60 per category).

\section{Human Evaluation Details}
\label{sec:app-userstudy}
The blind pairwise study involved 48 participants from our college: 35 from Computer Science, 7 from Data Science, 2 from Artificial Intelligence, and 4 graduate students or students from other programs.
Among them, 37 completed the Korean questionnaire and 11 completed the English version.
No personally identifiable information was collected during the study.
All participants evaluated the same ten questions regardless of their own department, comparing responses generated under the corresponding student profile.
For each of the 10 questions, participants were shown an A/B pair comparing \proposed with dense retrieval-based RAG under the \emph{deployment} setting.
To mitigate presentation-order bias, the positions of the \proposed and baseline responses in each A/B pair were counterbalanced across participants.
Table~\ref{tab:userstudy} presents the per-question results.

\begin{table}[t]
\centering\footnotesize
\setlength{\tabcolsep}{4pt}
\renewcommand{\arraystretch}{0.95}
\resizebox{\linewidth}{!}{
\begin{tabular}{l l ccc c}
\toprule
\textbf{ID} & \textbf{Query Topic} & Win & Tie & Lose & Win\% \\
\midrule
DS-01 & Graduation requirements       & 40 & 6  & 2  & 95 \\
DS-02 & Available major courses       & 25 & 4  & 18 & 58 \\
DS-03 & Required major courses        & 42 & 3  & 3  & 93 \\
AI-01 & Graduation requirements       & 43 & 3  & 2  & 96 \\
AI-02 & GE-required courses           & 27 & 13 & 7  & 79 \\
AI-03 & Foundational courses          & 21 & 15 & 11 & 66 \\
AI-04 & Required major courses        & 30 & 4  & 13 & 70 \\
CS-01 & Recommend major courses       & 43 & 2  & 1  & 98 \\
CS-02 & Graduation requirements       & 40 & 3  & 3  & 93 \\
CS-03 & Credits to graduate           & 43 & 2  & 1  & 98 \\
\midrule
\multicolumn{2}{l}{Total} & 354 & 55 & 61 & \textbf{85.3} \\
\bottomrule
\end{tabular}}
\caption{Results of the per-question human preference evaluation.}
\label{tab:userstudy}
\end{table}

\end{document}